\documentclass[prl,aps,tightenlines,superscriptaddress,floatfix,twocolumn,reprint]{revtex4-1} 
\usepackage{graphicx}
\usepackage{amsmath}
\usepackage{amsfonts,amsbsy}
\usepackage{amssymb}

\def\qs{Q_S}
\def\qp{ {\bf q}_T } 
\def\pp{ {\bf p}_T } 
\def\kp{ {\bf k}_T } 

\newcommand{\ud}{\mathrm{d}}

\begin{document}

\title{Azimuthal collimation of long range rapidity correlations by strong color fields \\
in high multiplicity hadron-hadron collisions}

\author{Kevin Dusling}
\affiliation{Physics Department, North Carolina State University, Raleigh, NC 27695, USA
}
\author{Raju Venugopalan}
\affiliation{Physics Department, Brookhaven National Laboratory,
  Upton, NY 11973, USA
}

\begin{abstract}
The azimuthal collimation of di-hadrons with large rapidity separations in high multiplicity p+p collisions at the LHC is described in the Color Glass Condensate  (CGC) effective theory~\cite{Dumitru:2010iy} by $N_c^2$ suppressed multi-ladder QCD diagrams that are enhanced $\alpha_S^{-8}$ due to gluon saturation in hadron wavefunctions. We show that quantitative computations in the CGC framework are in good agreement with data from the CMS experiment on per trigger di-hadron yields and  predict further systematics of these yields with varying trigger $p_T$ and charged hadron  multiplicity. Radial flow generated by re-scattering is strongly limited by the structure of the p+p di-hadron correlations.  In contrast, radial flow explains the systematics of identical measurements in heavy ion collisions. 

\end{abstract}

\maketitle

 The discovery of di-hadron correlations in high multiplicity proton-proton collisions~\cite{Khachatryan:2010gv}, long range in the angular (pseudo-rapidity) separation of the pairs relative to the beam axis and collimated in their relative azimuthal angle about this axis, provides significant insight into rare parton configurations in the proton and their dynamics in  hadronic collisions.

High multiplicity proton-proton collisions select ``hot spot" configuations of  wee gluon states in each proton.  Quantum Chromodynamics (QCD) predicts that such hot spots have a maximum occupancy of order $\alpha_S^{-1}$~\cite{Gribov:1984tu,Mueller:1985wy} ($\alpha_S$ being the QCD fine structure constant), and have a typical size $\sim 1/\qs$, where $\qs$ is a dynamical saturation scale. This scale grows with the energy and centrality of the collision;  when $\qs \gg \Lambda_{\rm QCD}$, the fundamental  QCD scale, highly occupied hadron wavefunctions can be described using weak coupling methods. 

A weak coupling effective field theory (EFT) that describes high density wee parton configurations in the proton is the Color Glass Condensate (CGC)~\cite{Gelis:2010nm}. When CGC's shatter in a high multiplicity collision, multi-particle production is a consequence of approximately boost invariant radiation from ``Glasma flux tubes" of transverse size $1/\qs$~\cite{Dumitru:2008wn}. Multiplicity distributions~\cite{Gelis:2009wh} derived from factorization theorems~\cite{Factorization} in this framework are in good agreement~\cite{Tribedy:2010ab,Tribedy:2011aa} with recent LHC data~\cite{Khachatryan:2010nk}. Long range rapidity correlations of gluons computed in the CGC EFT~\cite{Dusling:2009ni} were previously shown to be in qualitative agreement~\cite{Dumitru:2010iy} with the CMS di-hadron correlation data. 

\begin{figure}[t]
\centering
\includegraphics[scale=.6]{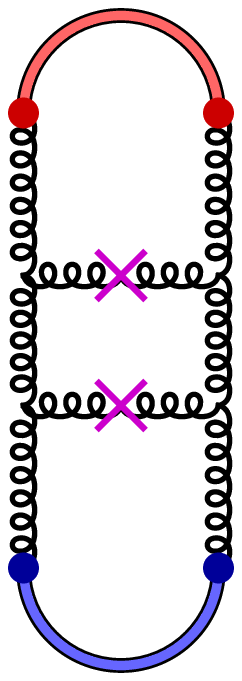}
\hspace{0.75in}
\includegraphics[scale=.6]{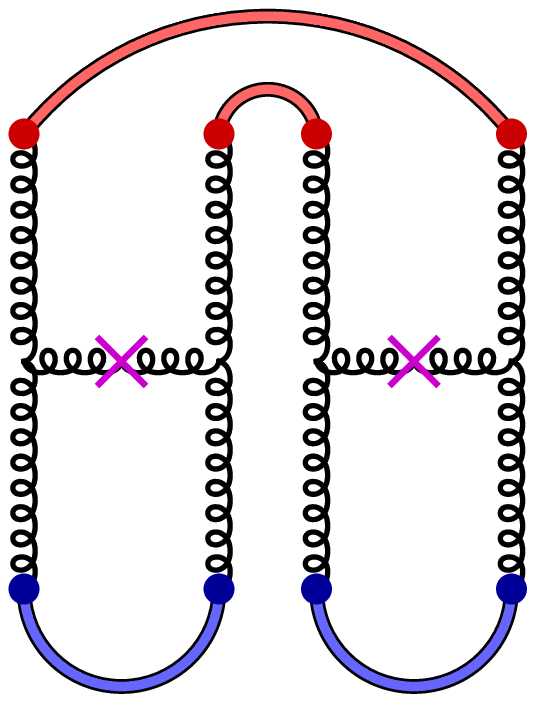}
\caption{Representative back-to-back (left) and Glasma graphs (right) in perturbative QCD.}
\label{fig:graph}
\end{figure}

A source of long range rapidity correlations in hadron-hadron collisions are back-to-back gluons emitted from a single $t$-channel gluon ladder; another source, called ``Glasma graphs" are  gluons emitted from two separate ladders.  Representative graphs of each are shown in fig.~(\ref{fig:graph}). In the ``dilute" high $p_T$ perturbative limit of QCD, the back-to-back contribution is dominant.  However, at high parton densities, when $\qs^2 \gg \Lambda_{\rm QCD}^2$, and $p_T^2\sim \qs^2$, the effective coupling of gluons in ladders to strong color sources at higher rapidities changes from $g \rightarrow 1/g$.  This corresponds to an enhancement of Glasma graphs by $\alpha_S^{-8}$ compared to the $\alpha_S^{-4}$ enhancement of the back--to--back graphs. Equally important are the very different azimuthal structures of the two long range rapidity correlations. Back-to-back graphs, as the name suggests, are kinematically constrained to be peaked ``away side" at relative azimuthal angle $\Delta \phi \sim \pi$ and have a negligible``near side" collimation at $\Delta \phi\sim 0$ as seen in high energy asymptotics that produce a long range rapidity correlation~\cite{Leonidov:1999nc,Fadin:1996zv}.  

In contrast, Glasma graphs give identical near and away side contributions
because gluon emission is from independent ladders. As also noticed
elsewhere~\cite{Bartels:2011qi,Levin:2011fb}, these correlated contributions
producing an azimuthal collimation are of order $1/N_c^2$; their contribution
would be negligible if one did not have the $\alpha_S^{-8}$ enhancement in the
high multiplicity region.  Within the CGC framework itself, there are leading
$N_c$ correlated multi-ladder contributions~\cite{Factorization} which are long
range in rapidity. However, these do not produce an azimuthal
collimation~\cite{Dumitru:2010mv,Dumitru:2011zz}. Likewise, there can be
pomeron loop effects outside the framework of ref.~\cite{Factorization}; again,
these either do not give a collimation or the collimation vanishes rapidly with
increasing rapidity~\cite{Kovner:2010xk,Kovner:2011pe}. However, a
Zero-Yield-at Minimum (ZYAM) procedure~\cite{CMS-PAS-HIN-11-006} used by the CMS collaboration to compute the per trigger near side di-hadron yield~\cite{Li:2011mp} only measures contributions that are collimated in  $\Delta\phi$ above the $\Delta\phi$-independent background. The ZYAM procedure allow one to focus on those di-hadron correlations that produced a collimated near side yield.  This is also fortuitous because the relative normalization between Glasma graphs and back-to-back graphs necessary to reproduce the di-hadron yield in the entire $\Delta \Phi$ range is not under theoretical control. 
%In this letter, we will focus on the quantitative contribution of the Glasma graphs and compare the systematics to the CMS proton-proton collision data. We will also explore the relative role of radial flow in generating near side yields in proton-proton and nucleus-nucleus collisions.  

%%%%%%%%%%%%%%%%%%%%%%%%%%%%%%%%%%%%%%%%%%%%%%%%%%%%%%%%%%%%%%%%%%%%%%%%%
%%%%%%%%%%%%%%%%%%%%%%%%%%%%%%%%%%%%%%%%%%%%%%%%%%%%%%%%%%%%%%%%%%%%%%%%% 

The correlated two gluon glasma distribution, expressed in terms of the two particle momentum space rapidities $y_{p,q}$ and transverse momenta $p_\perp,q_\perp$, is  ~\cite{Dusling:2009ni} 
 \begin{align}
\frac{d^2N_{\rm \sl Glasma}^{\rm \sl corr.}}{d^2\pp d^2\qp dy_p dy_q}
= \frac{C_2}{\pp^2\qp^2} \int_{\kp} (D_1 + D_2) \, ,
\label{eq:Glasma-corr}
\end{align}
where~\footnote{This prefactor corrects for a typo and is hence a factor of 4 larger than in refs.~\cite{Dusling:2009ni,Dumitru:2010iy} but in agreement in the appropriate limit with the corresponding expression in ref.~\cite{Gelis:2009wh}.}  $C_2 = \frac{\alpha_S^{2}}{4\pi^{10}}\frac{N_C^2 S_\perp}{(N_C^2-1)^3 \zeta}$ and
\begin{align}
D_1 &= \Phi_{A_1}^2(y_p,\kp)\Phi_{A_2}(y_p,\pp-\kp)
D_{A_2}\nonumber \\
D_2 &= \Phi_{A_2}^2(y_q, \kp)\Phi_{A_1}(y_p,\pp-\kp)
D_{A_1}\, ,
\end{align}
with  $D_{A_{2(1)}} = \Phi_{A_{2(1)}}(y_q,\qp+\kp)+\Phi_{A_{2(1)}}(y_q,\qp-\kp)$. 
For our computation, we will also need the single inclusive gluon distribution
\begin{align}
\frac{\ud N_1}{\ud y_p\ud^2\pp }
=\frac{C_1}{\pp^2}
\int_{\kp}\!\!\!
\Phi_{A_1}(y_p,\kp)\Phi_{A_2}(y_p,\pp-\kp)\,,
\label{eq:single-incl}
\end{align}
with the coefficient $C_1=\frac{\alpha_s N_C S_\perp}{4\pi^6 (N_C^2-1)}$. 
The important ingredient in these expressions is the 
unintegrated gluon distribution (UGD) per unit transverse area, defined as 
\begin{equation}
\Phi_A(y,k_\perp) = {\pi N_C k_\perp^2\over 2\alpha_S}\int_0^\infty dr_\perp r_\perp J_0(k_\perp r_\perp)  [1-{\cal T}_A(y,r_\perp)]^2\, 
\label{eq:unint-gluon}
\end{equation}
Here ${\cal T}_A$ is the forward scattering amplitude of a quark-antiquark dipole of transverse 
size $r_\perp$ on the target $A$; it, or equivalently, the UGD, is a universal quantity that can be determined by solving the Balitsky-Kovchegov (BK) 
equation~\cite{Balitsky:1995ub,Kovchegov:1999yj} as a function of the rapidity $y=\log\left(x_0/x\right)$.  Solutions of the running coupling BK (rcBK) equation \footnote{The BK equation is valid at large $N_c$ for dense color sources as a limit of the Balitsky-JIMWLK hierarchy~\cite{Gelis:2010nm}. The finite $N_c$ corrections are however at most $1/N_c^2$; in practice, corrections to the evolution are 
less than a percent~\cite{Dumitru:2011vk,Rummukainen:2003ns}.} used to compute structure functions are in good agreement with the HERA inclusive data~\cite{Albacete:2009fh}. 

The eqs.~(\ref{eq:Glasma-corr}) and (\ref{eq:single-incl}) are obtained from the CGC formalism in ref.~\cite{Factorization} for collisions of high high parton density sources, as may be realized in nucleus-nucleus and high multiplicity proton-proton collisions. We emphasize that, albeit not shown explicitly in fig.~(\ref{fig:graph}), the derivation of eq.~(\ref{eq:Glasma-corr}) in ref.~\cite{Dusling:2009ni} encodes the effect of radiation between the sources and the triggered gluons as well as the radiation between the gluons. In obtaining these results, the distribution of color sources is assumed to be a non-local Gaussian distribution, whose variance is simply related to $\Phi_A(y,k_\perp)$. This assumed distribution has been shown recently to provide a good approximate solution to the Balitsky-JIMWLK hierarchy for $n$-point lightlike Wilson line correlators~\cite{Dumitru:2011vk,Iancu:2011nj}. The unintegrated gluon distribution in eq.~(\ref{eq:unint-gluon}) have a ``bell-shaped" structure peaked at $\qs$, with the peak moving to larger $k_\perp$ with increasing rapidity. Thus eqs.~(\ref{eq:Glasma-corr}) and (\ref{eq:single-incl}) are infrared finite unlike the expressions in ref.~\cite{Gelis:2009wh}. However, like the latter, they do not include multiple scattering contributions that are present in the formalism of ref.~\cite{Factorization} and contribute for $k_\perp \leq \qs$. Their effect on eq.~(\ref{eq:Glasma-corr}) is given by a non-perturbative constant\footnote{Multiple scattering effects will quantitatively affect but not qualitatively alter the $\phi$ distributions in the infrared. Note also, to avoid confusion, $\zeta$ here plays the same role as the constant $\kappa$ in ref.~\cite{Gelis:2009wh}.} $\zeta$ estimated numerically in ref.~\cite{Lappi:2009xa} to be in the range $1/3$--$3/2$. Fits to p+p multiplicity distributions for a range of energies at the LHC and A+A multiplicity distributions at RHIC obtained $\zeta=1/6$~\cite{Tribedy:2010ab,Tribedy:2011aa}. Given uncertainties in the numerical computation we will use the latter empirical value instead. 

Qualitatively, the origin of a large collimated contribution from Glasma graphs
occurs because the integrand in eq.~(\ref{eq:Glasma-corr}) is large when the
peaks of the ``bell-shaped" unintegrated distributions strongly overlap, $|\kp|
\sim |\pp - \kp| \sim |\qp \pm \kp| \sim \qs$, giving a collimation at $\Delta\Phi=0,\pi$. In practice, the result is smeared by fragmentation effects as well as the details of the integration. We will therefore in the rest of this letter focus on the quantitative contribution of the Glasma graphs and compare the systematics to the CMS proton-proton collision data. We will also explore the relative role of radial flow in generating near side yields in proton-proton and nucleus-nucleus collisions.  

The initial condition in rcBK is the McLerran-Venugopalan-like ~(MV)- initial condition~\cite{MV} for ${\cal T}_A$ at the the initial rapidity (corresponding to Bjorken $x\equiv x_0 = 0.01$). The minimum bias saturation scale  $Q_0^2$  in the MV initial condition at the initial rapidity and the  transverse area $S_\perp$ are adjusted to reproduce the single inclusive p+p distribution in eq.~(\ref{eq:single-incl})--for a more detailed discussion, see ref.~\cite{Tribedy:2010ab}. Diffractive scattering results from HERA indicate a strong dependence of the saturation scale on impact parameter or the centrality of the collision~\cite{Kowalski:2006hc}. The centrality dependence of eq.~(\ref{eq:unint-gluon}) is therefore  studied here by keeping $S_\perp$ fixed and varying $Q_0^2$ at the initial rapidity scale~\footnote{Since the di-hadron yields are normalized per trigger, the $S_\perp$ dependence drops out for the quantities studied here.}.  We follow the   results of the HERA studies in ref.~\cite{Kowalski:2006hc} and choose $Q_0^2$ values in the fundamental representation of $0.15$, $0.3$, $0.45$ and $0.6$ GeV$^2$ to represent different centralities in computing the di-hadron yield~\footnote{These initial values should not be confused with the saturation momentum, defined here as the peak of $\phi(y,k_\perp)$, at the much smaller values of $x\sqrt{s}=p_T e^{\pm y}$ probed in the di-hadron studies here, which span (in the adjoint representation), $\qs^2\left(x\sim 10^{-4\div 5}\right)\sim 1-4$ GeV$^2$ respectively for initial saturation scales used in this work.}.

Further calculational details are as follows. The strong coupling constant, $\alpha_S$, is evaluated at the saturation scale.  Because the di-hadrons of interest are widely separated in rapidity, we assume that gluons fragment independently. Fragmentation functions at forward rapidities are not particularly well known~\cite{Sassot:2010bh}; these will be better constrained as more data from the LHC becomes available. For our purposes, we consider a soft fragmentation function $D_1(x) = 3 (1-x)^2/x$ and a hard one 
$D_2(x) = 2 (1-x)/x$; the former is closer to the NLO fit function for gluon
fragmentation to pions~\cite{Kniehl:2000fe}, while the latter allows for hadrons
to carry on average a larger fraction of the gluon momentum. Finally, we
introduce an overall constant $K$ factor which is the only parameter fit to the
data presented here. It accounts for higher order corrections (both in our
framework as well as in the fragmentation functions) in addition to corrections
necessary to fully implement the experimental acceptance~\footnote{One example
of the latter is $N_{\rm trig}$, which is required experimentally to be $\geq
2$. Properly implementing this constraint would require Monte--Carlo simulation
and is beyond the scope of this work.}.

For our analysis of the CMS data, we define 
\begin{align}
N_{\rm trig}=\int_{-2.4}^{+2.4}\! \!d\eta \!\!\int_{p_T^{\rm min}}^{p_T^{\rm max}}\!\!\!\! d^2\pp\!\!\int_{0}^1 \!\!  dz \frac{D(z)}{z^2} \frac{dN}{d\eta \,d^2 \pp}\left(\frac{p_{\textrm{T}}}{z}\right)
\label{eq:ntrig}
\end{align}
and 
\begin{align}
&\frac{d^2N}{d\Delta \phi} = K \int_{-2.4}^{+2.4} \!d\eta_p  \,d\eta_q \,\, {\cal A}\left(\eta_p,\eta_q\right) \\
&\!\!\times\int_{p_T^{\rm min}}^{p_T^{\rm max}} \frac{dp_T^2}{2} \int_{q_T^{\rm min}}^{q_T^{\rm max}}\frac{ d q_T^2}{2}\;\int d\phi_p \int d\phi_q\; \delta\left(\phi_p-\phi_q-\Delta\phi\right) \nonumber\\
&\!\!\times \int_{0}^1\!\! dz_1 dz_2 \frac{D(z_1)}{z_1^2}\, \frac{D(z_2)}{z_2^2}
 \frac{d^2N_{\rm \sl Glasma}^{\rm \sl corr.}}{d^2\pp d^2\qp d\eta_p d\eta_q}\left(\frac{p_{\textrm{T}}}{z_1},\frac{q_{\textrm{T}}}{z_2},\Delta\phi \right)\nonumber
\label{eq:dihadron}
\end{align}
Here $p_T^{\rm min (max)}$ and $q_T^{\rm min (max)}$ refer to bounds on the range of the trigger and associated hadron momenta respectively. Likewise, 
$\Delta\eta_{\rm min}(\Delta \eta_{max})=2.0 (4.0)$ denote the pseudo-rapidity gap~\footnote{We replace the rapidity $y$ with the pseudo-rapidity $\eta$ 
which is a good approximation for the $p_T$, $q_T$ of interest.} of hadrons within the experimental acceptance ${\cal A}\left(\eta_p,\eta_q\right)\equiv \theta\left( \vert\eta_p -\eta_q\vert - \Delta\eta_{min}\right)\, \theta\left(\Delta\eta_{\max} - \vert \eta_p-\eta_q\vert\right)$. 

%%%%%%%%%%%%%%%%%%%%%%%%%%%%%%%%%%%%%%%%%%%%%%%%%%%%%%%%%%%%%%%%%%%%%%%%%
\begin{figure}[t]
\centering
\includegraphics[scale=0.8]{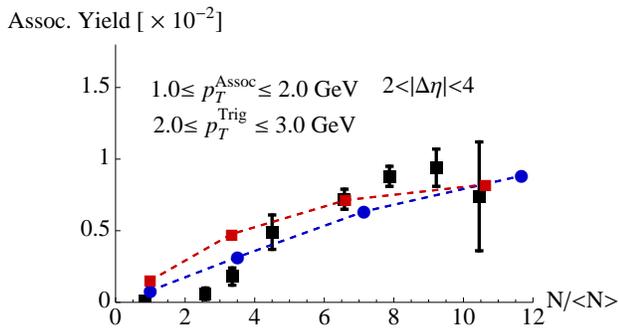}
\caption{Associated yield for four different initial saturation scales representing different centralities. The blue circles (red squares) represent 
the softer (harder) $D_1(z)$ ($D_2(z))$ fragmentation functions.  Dashed lines connect computed points to guide the eye.  Data points are from the CMS collaboration \cite{CMS-PAS-HIN-11-006}.}
\label{fig:cent1}
\end{figure}
%%%%%%%%%%%%%%%%%%%%%%%%%%%%%%%%%%%%%%%%%%%%%%%%%%%%%%%%%%%%%%%%%%%%%%%%% 

The strength of the correlation in $\Delta \phi$ is quantified by the associated yield computed using the ZYAM procedure defined to be 
\begin{align}
\!\textrm{Assoc. Yield} = \frac{1}{N_{\rm trig}}\int_0^{\Delta\phi_{\rm min.}} \!\!\!\! d\Delta\phi \frac{d^2N}{d\Delta\phi}-\left.\frac{d^2N}{d\Delta\phi}\right|_{\Delta\phi_{\rm min.}}\!\!\!\!\!
\end{align}
where $\Delta\phi_{\rm min.}$ is the angle at which the two particle correlation strength is minimal.  In fig.~(\ref{fig:cent1}), we plot the associated yield as a function of charged particle multiplicity, per minimum bias multiplicity, for the stated windows in $\Delta \eta$ and in the associated and trigger particle transverse momenta. As noted previously, the charged particle multiplicity is varied by changing $Q_0^2$ in the initial conditions for rcBK evolution. We see that the agreement is quite good, especially at the highest multiplicities where we expect our formalism to perform best.  At lower multiplicities, the effect of high order corrections as well as impact parameter fluctuations become more important. 

\begin{figure}[t]
\centering
\includegraphics[scale=0.8]{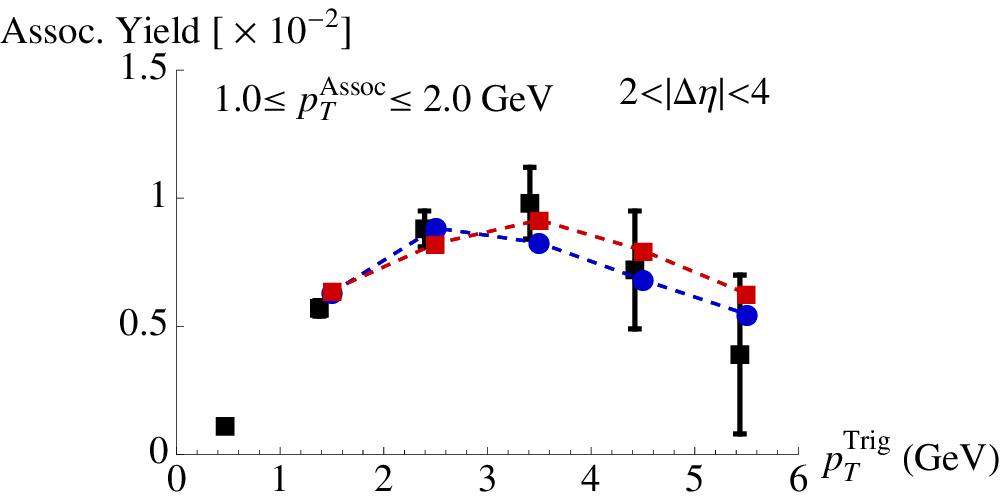}
\includegraphics[scale=0.8]{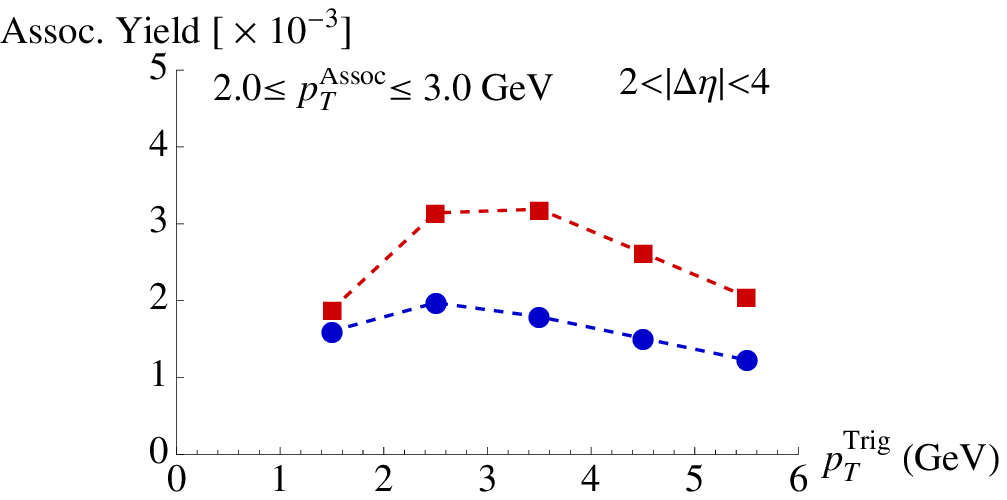}
\includegraphics[scale=0.8]{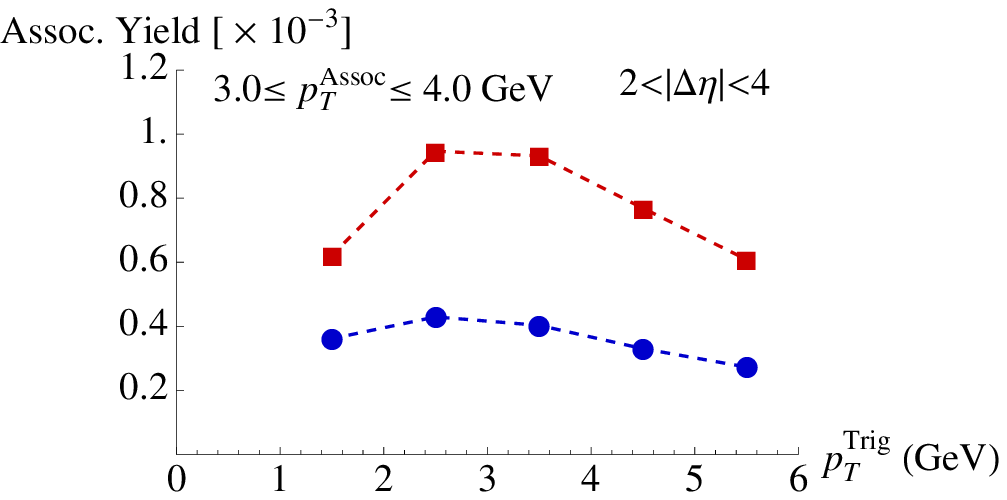}
\caption{Associated yield for central $p+p$ ($Q_0^2=0.6$ GeV$^2$) collisions using soft (hard) $D_1(z)$-blue circles ($D_2(z)$-red squares) fragmentation functions. Dashed lines connect computed points to guide the eye.  The black squares are the vailable CMS data \cite{CMS-PAS-HIN-11-006} for the $N\geq 110$ multiplicity bin of $pp$ collisions at $\sqrt{s}=7$ TeV.  The middle and bottom figures are predictions for the labeled associated $p_T$ windows.}
\label{fig:pty}
\end{figure}

In fig.~(\ref{fig:pty}), we plot the associated yield versus the $p_T$ trigger window for three associated particle windows as labeled.  The top figure corresponding to $1.0\leq p_T^{Assoc} \leq 2.0$ GeV also shows the recent CMS measurements of the same quantity.  Even though the overall normalization of our calculation needed to be augmented by a constant K-factor, $K=2.3$,  the momentum dependence of our results is parameter free.  The other two plots are absolute predictions; though, as shown, they are quite sensitive to fragmentation. The sensitivity of the associated yield to different momentum cuts in our calculation stems from an intrinsic scale (the saturation momentum) where the initial state wave--function is peaked.  As argued in \cite{Dumitru:2010iy} maximal angular correlations occur when $\left| \pp \right| \sim \left| \qp \right|\sim \qs$ and when $\pp$ and $\qp$ are parallel.  This signal persists after including fragmentation and shows good agreement with the measured high multiplicity pp data. Our model computation provides strong support to the qualitative idea that a significant near side angular correlation at semi-hard trigger and associate hadron momenta of $2-4$ GeV is evidence of saturation dynamics. 
%The angular correlation data clearly demonstrate a preference for a momentum scale on the order of 2--4 GeV and our model calculations lend support that this scale is indeed related to saturation.

%%%%%%%%%%%%%%%%%%%%%%%%%%%%%%%%%%%%%%%%%%%%%%%%%%%%%%%%%%%%%%%%%%%%%%%%%
%%%%%%%%%%%%%%%%%%%%%%%%%%%%%%%%%%%%%%%%%%%%%%%%%%%%%%%%%%%%%%%%%%%%%%%%% 

\begin{figure}[t]
\centering
\includegraphics[scale=0.8]{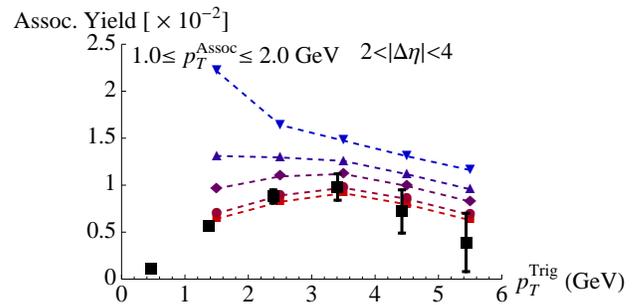}
\caption{Effect of transverse flow on the intrinsic $pp$ correlation using the hard $D_2(z)$ fragmentation function.  Boosts from bottom to top: 
$\beta=0,0.1,0.2,0.25,0.3$.}
\label{fig:ppFlow}
\end{figure}

Because the number of particles produced in the highest multiplicity $pp$ collisions are comparable to those in {\it Cu Cu} collisions one may speculate that (above and beyond the collimation provided by our intrinsic QCD effect) collective flow contributes significantly to the angular correlation~\cite{Bozek:2010pb,Werner:2010ss}. To test this hypothesis, we employ a radial boost model where the angular distribution in $\Delta\phi$ in the laboratory frame is related to the corresponding distribution in $\Delta\tilde\phi$ in the local rest frame,
\begin{align}
\frac{d^2N}{d\Delta\phi}=\int_{-\pi}^\pi d\Psi \mathcal{J}\left(\Psi,\Delta\phi\right) \frac{d^2N}{d\Delta\tilde{\phi}}\left(\Delta\tilde{\phi}\left(\Psi,\Delta\phi\right)\right)\, ,
\end{align}
where $\mathcal{J}$ is the Jacobian~\footnote{The relationship between the opening angles $\Delta\phi$ in the laboratory frame and the opening angle $\Delta\tilde\phi$ in the local rest frame under a Lorentz boost having velocity $v=\beta c$ and direction $2\Psi=\phi_p+\phi_q$ is given by
\begin{align*}
&2\sin^2\left(\frac{\Delta\tilde{\phi}}{2}\right)=\\
&\frac{\sqrt{1-\beta^2}\left(1-\cos\left(\Delta\phi\right)\right)}{1-2\beta\cos\Psi                                   \cos\left(\frac{\Delta\phi}{2}\right)+                                          \frac{\beta^2}{2}\left(\cos\left(\Delta\phi\right)+ \cos\left(2\Psi\right)\right)}\;.
\end{align*}
In addition one must include a Jacobian factor 
\begin{align*}
\mathcal{J}=\frac{1-\beta^2}{\left(1-\beta\cos\left(\Psi+\Delta\phi/2\right)\right)\left(1-\beta\cos\left(\Psi-\Delta\phi/2\right)\right)}\;,
\end{align*} 
when changing between the local rest frame and laboratory distributions. 
} relating distributions in the two frames. As transverse flow further collimates the signal, the overall strength of the associated yield will increase. However, the momentum dependence changes as well.
% as the boosted signal still depends on the intrinsic angular correlation that is present in the local rest frame.  
The effect of the boost is demonstrated in fig.~(\ref{fig:ppFlow}). Starting with our correlation in the local rest frame (red squares), we show the result after transverse boosts of 
(bottom to top) $\beta=0.1,0.2, 0.25,0.3$. One notices a qualitative change in the shape of the associated yield versus $p_T^{\rm trig}$. For smaller transverse boosts the dependence on $p_T^{\rm trig}$ is given by the intrinsic angular correlation generated by the Glasma graph of fig.~(\ref{fig:graph}).  For large boost velocities the associated yield is driven by the collimation of the $\Delta\phi$ independent pedestal computed from the same graph.  Without any transverse flow this pedestal (or underlying event) is removed by the ZYAM procedure and therefore does not contribute to the associated yield.  But after collimation, its signal exceeds that of the intrinsic angular correlation.  The change in shape therefore suggests an upper bound of $\beta =0.25$ in this simple model of flow in {\it pp} collisions.

%%%%%%%%%%%%%%%%%%%%%%%%%%%%%%%%%%%%%%%%%%%%%%%%%%%%%%%%%%%%%%%%%%%%%%%%%
%%%%%%%%%%%%%%%%%%%%%%%%%%%%%%%%%%%%%%%%%%%%%%%%%%%%%%%%%%%%%%%%%%%%%%%%% 

\begin{figure}[t]
\centering
\includegraphics[scale=0.8]{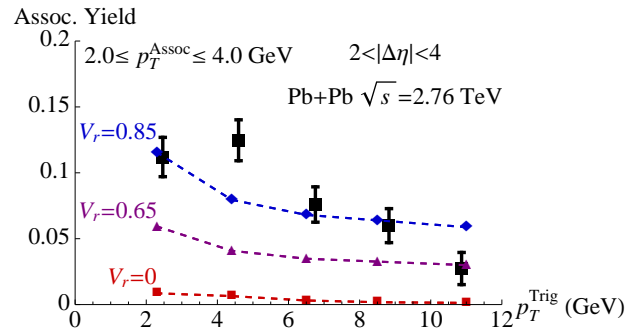}
\caption{Computations of the associated per trigger yield in {\it Pb Pb} collisions at $\sqrt{s}=2.76$ TeV using $Q_0^2=0.9$ GeV$^2$ in the fundamental representation compared to the CMS data \cite{CMS-PAS-HIN-11-006}.  The curves shown, with the $D_2(z)$ fragmentation function, are for transverse boosts of $\beta=0, 0.65, 0.85$.  At large flow velocities, the intrinsic angular correlation is entirely washed out.}
\label{fig:PbPb}
\end{figure}

This is in complete contrast to heavy ion collisions where we expect flow to dominate the angular correlation~\cite{Voloshin:2003ud,Pruneau:2007ua,Dumitru:2008wn}.  We demonstrate this with a comparison of the $p_T^{\rm trig}$ dependence of the (collimated by flow) pedestal in the Glasma with data from {\it Pb Pb} collisions at $\sqrt{s}=2.76$ TeV in fig.~(\ref{fig:PbPb}). The agreement is quite good considering the very simple model of radial flow considered here. Flow effects here completely dwarf the intrinsic QCD correlations that were the dominant effect generating the near side azimuthal collimation in  {\it pp} collisions.  We should stress however that the pedestal (while independent of $\Delta\phi$ in the local rest frame) is also an intrinsic two--particle correlation generated by the Glasma graph and the $p_T^{\rm trig}$ dependence seen in fig.~(\ref{fig:PbPb}) is representative of this underlying dynamics.
%%%%%%%%%%%%%%%%%%%%%%%%%%%%%%%%%%%%%%%%%%%%%%%%%%%%%%%%%%%%%%%%%%%%%%%%%
%%%%%%%%%%%%%%%%%%%%%%%%%%%%%%%%%%%%%%%%%%%%%%%%%%%%%%%%%%%%%%%%%%%%%%%%% 

\section*{Acknowledgements}
We thank Wei Li for helpful clarifications regarding the CMS results. K.D.  and  R.V are  supported by the US Department of Energy under DOE Contract Nos.
DE-FG02-03ER41260 and DE-AC02-98CH10886 respectively.

\end{document}